\documentclass[11pt]{article}
\usepackage{epsfig}

\def\la{\langle}\def\ra{\rangle}
\def\be{\begin{eqnarray}}
\def\ee{\end{eqnarray}}
\def\lsim{\mathrel{\rlap{\lower3pt\hbox{\hskip1pt$\sim$}}
     \raise1pt\hbox{$<$}}} 
\def\gsim{\mathrel{\rlap{\lower3pt\hbox{\hskip1pt$\sim$}}
     \raise1pt\hbox{$>$}}} 

\begin{document}
\begin{titlepage}
\hfill {KIAS-P03073}

\hfill {\today: hep-ph/}

\begin{center}
\ \\
{\Large \bf  Vector Mesons and Dense Skyrmion Matter}
\\
\vspace{.30cm}

Byung-Yoon Park$^{a,b}$, Mannque Rho$^{a,c}$
\\ and Vicente Vento$^{a,d}$

\vskip 0.20cm

{(a) \it School of Physics, Korea Institute for Advanced Study,
Seoul 130-772, Korea}

{(b) \it Department of Physics,
Chungnam National University, Daejon 305-764, Korea}\\
({\small E-mail: bypark@chaosphys.cnu.ac.kr})

{(c) \it Service de Physique Th\'eorique, CE Saclay}\\
{\it 91191 Gif-sur-Yvette, France}\\
({\small E-mail: rho@spht.saclay.cea.fr})

{(d) \it Departament de Fisica Te\`orica and Institut de
F\'{\i}sica
Corpuscular}\\
{\it Universitat de Val\`encia and Consejo Superior
de Investigaciones Cient\'{\i}ficas}\\
{\it E-46100 Burjassot (Val\`encia), Spain} \\ ({\small E-mail:
Vicente.Vento@uv.es})

\end{center}
\vskip 0.3cm

\centerline{\bf Abstract}

In our continuing effort to understand hadronic matter at high
density, we have developed a unified field theoretic formalism for
dense skyrmion matter using a single Lagrangian to describe
simultaneously both matter and meson fluctuations and studied {\it
in-medium} properties of hadrons. Dropping the quartic Skyrme
term, we incorporate into our previous Lagrangian the vector
mesons $\rho$ and $\omega$ in a form which is consistent with the
symmetries of QCD. The results that we have obtained, reported
here, expose a hitherto unsuspected puzzle associated with the
role the $\omega$ meson plays at short distance. Since the
$\omega$ meson couples to baryon density, it leads to a pseudo-gap
scenario for the chiral symmetry phase transition, which is at
variance with standard scenario of QCD at the phase transition. We
find that in the presence of the $\omega$ mesons, the
scale-anomaly dilaton field is prevented from developing a
vanishing vacuum expectation value at the chiral restoration, as a
consequence of which the in-medium pion decay constant does not
vanish. This seems to indicate that the $\omega$ degree of freedom
obstructs the ``vector manifestation" which is considered to be a
generic feature of effective field theories matched to QCD.

\vskip 0.5cm

\vskip 0.3cm \leftline{Pacs: 12.39-x, 13.60.Hb, 14.65-q, 14.70Dj}
\leftline{Keywords: skyrmion, dilaton, vector mesons, dense
matter, BR scaling}

\end{titlepage}
\section{Introduction}

The recent experimental and theoretical developments in dense
matter physics have shown that the phase diagram of hadronic
matter is rich and highly non-trivial. At high temperature and/or
density, hadrons have quite different properties than at normal
conditions. Chiral symmetry is believed to be restored and
therefore the quark condensate $\la\bar{q} q\ra$ of QCD, its order
parameter, is expected to drop as matter is heated and/or
compressed.

In trying to understand what happens to hadrons under extreme
conditions, it is necessary that the theory adopted for the
description be consistent with QCD. In terms of effective theories
this means that they should match to QCD at a scale close to the
chiral scale $\Lambda_\chi \sim 4\pi f_\pi \sim 1$ GeV. It has
been shown that this matching can be effectuated in the framework
of hidden local symmetry and leads to what is called ``vector
manifestation (VM)"\cite{harada,HY:PR} which provides a
theoretical support to the in-medium behavior of hadrons predicted
in 1991\cite{br,BR03}.

We have carried out several studies in addressing the problem of
dense matter within the general scheme we are adopting. We have
shown that the simple skyrmion Lagrangian can describe both
infinite matter and pionic fluctuations thereon and their behavior
as the density increases \cite{skyrmionmatter}. However in this
first study we found a puzzling feature, namely that the Wigner
phase represented by half-skyrmion matter supported a
non-vanishing pion decay constant. This was interpreted as a
possible signal for a pseudo-gap scenario which according to
\cite{harada} would be at variance with QCD and therefore could
not be realized in nature.

In \cite{sliding,pionvel}, we incorporated into the standard
skyrmion model the scale anomaly of QCD. In fact it was this
feature which led to BR-scaling in \cite{br} and therefore we were
able to reformulate this phenomenon in a more accurate way. In
this scenario $f_\pi$ vanishes at the chiral phase transition and
other properties of the vector manifestation scenario are
reproduced. Thus large $N_c$ and scale anomaly seem to be the
minimal ingredients to satisfy the matching between the effective
theories and QCD.

The purpose of the present paper is to do away with the ad hoc
Skyrme quartic term and extend the model by incorporating the
lowest-lying vector mesons, namely the $\rho$ and the $\omega$. It
is known that these vector mesons play a crucial role in
stabilizing the single nucleon system \cite{zahedbrown} as well as
in the saturation of normal nuclear matter~\cite{walecka} and
moreover it would be important to determine how their properties
behave in the medium, since they lead to measurable quantities
that can be confronted by different descriptions.

We consider a skyrmion-type Lagrangian with vector mesons
possessing hidden local gauge symmetry
\cite{zahedbrown,ellis,bando}, spontaneously broken chiral
symmetry and scale symmetry. Such a theory might be considered as
a better approximation to reality than the extreme large $N_c$
approximation to QCD represented by the Skyrme model. In our
approach a skyrmion description for the multi-baryon system
(including infinite matter) can be obtained with the parameters of
the theory fixed by meson dynamics. Our interest here lies in the
investigation of how the vector mesons affect the description of
nuclear matter, i.e., the density and the properties of the chiral
restoration phase transition. For this purpose we look at the
ground state of the infinite matter system which is a soliton
solution (i.e., skyrmion matter) of the Lagrangian. As one varies
the density of the system the parameters of the theory involving
this process must adapt to the density of the skyrmion background.
We will describe their change with density in our model for
nuclear matter.

The content of this paper is as follows. In Section 2, we write
down our model Lagrangian which is the simplest form of skyrmion
Lagrangian, which incorporates the vector mesons in a way
consistent with the scale anomaly and symmetries of QCD . In
Section 3, a single skyrmion is analyzed to define the parameters
of the theory at zero density. In Section 4 we analyze the two
skyrmion case in order to assess the behavior of the vector-meson
components of the $B \ne 1$ skyrmion. The low density realization
of the skyrmion matter in our model, an FCC crystal, is described
in Section 5. Section 6 is devoted to the study of the chiral
restoration phase transition and the important role of the
$\omega$ in the model. Some concluding remarks are given in
Section 7.

\section{Model Lagrangian}
The starting point of our work is the skyrmion Lagrangian
introduced in our previous work \cite{sliding} which contains the
proper, but minimal, realization of spontaneous symmetry and scale
symmetry breaking, fundamental properties of QCD, in the effective
mesonic degrees of freedom. To it we incorporate the vector
mesons, $\rho$ and $\omega$, maintaining the adequate
implementation of the symmetry realizations. Specifically, the
model Lagrangian, which we investigate, is given by \cite{MRY00}

\begin{eqnarray}
{\cal L} &=& \frac{f_\pi^2}{4} \left(\frac{\chi}{f_\chi}\right)^2
\mbox{Tr}(\partial_\mu U^\dagger \partial^\mu U) + \frac{f_\pi^2
m_\pi^2}{4} \left(\frac{\chi}{f_\chi}\right)^3
    \mbox{Tr}(U+U^\dagger-2)
\nonumber\\
&&
-\frac{f_\pi^2}{4} a \left(\frac{\chi}{f_\chi}\right)^2
 \mbox{Tr}[\ell_\mu + r_\mu + i(g/2)
( \vec{\tau}\cdot\vec{\rho}_\mu + \omega_\mu)]^2
-\textstyle \frac{1}{4} \displaystyle
\vec{\rho}_{\mu\nu} \cdot \vec{\rho}^{\mu\nu}
-\textstyle \frac{1}{4}  \omega_{\mu\nu} \omega^{\mu\nu}
\nonumber\\
&& +\textstyle\frac{3}{2} g \omega_\mu B^\mu
+\textstyle\frac{1}{2} \partial_\mu \chi \partial^\mu \chi
-\displaystyle \frac{m_\chi^2 f_\chi^2}{4} \left[ (\chi/f_\chi)^4
(\mbox{ln}(\chi/f_\chi)-\textstyle\frac14) + \frac14 \right],
\label{lag}\end{eqnarray}
where

$$
U =\exp(i\vec{\tau}\cdot\vec{\pi}/f_\pi) \equiv \xi^2,
\eqno(\mbox{\ref{lag}.a})$$
$$
\ell_\mu = \xi^\dagger \partial_\mu \xi, \mbox { and }
r_\mu = \xi \partial_\mu \xi^\dagger,
\eqno(\mbox{\ref{lag}.b})$$
$$
\vec{\rho}_{\mu\nu} = \partial_\mu \vec{\rho}_\nu
- \partial_\nu \vec{\rho}_\mu + g \vec{\rho}_\mu \times \vec{\rho}_\nu,
\eqno(\mbox{\ref{lag}.c})$$
$$
\omega_{\mu\nu}=\partial_\mu\omega_\nu-\partial_\nu\omega_\mu,
\eqno(\mbox{\ref{lag}.d})$$
$$
B^\mu =  \frac{1}{24\pi^2} \varepsilon^{\mu\nu\alpha\beta}
\mbox{Tr}(U^\dagger\partial_\nu U U^\dagger\partial_\alpha U
U^\dagger\partial_\beta U).
\eqno(\mbox{\ref{lag}.e})$$
Note that the Skyrme quartic term is not present in the model. The
vector mesons, $\rho$ and $\omega$, are incorporated as dynamical
gauge bosons for the local hidden gauge symmetry of the non-linear
sigma model Lagrangian and the dilaton field $\chi$ is introduced
so that the Lagrangian has the same scaling behavior as QCD. The
physical parameters appearing in the Lagrangian are summarized in
Table. 1. Throughout this paper, we take the following convention
for the indices: (i) $a,b,\cdots =1,2,3$ (Euclidean metric) for
the isovector fields; (ii) $i,j,\cdots=1,2,3$ (Euclidean metric)
for the spatial components of normal vectors;  (iii)
$\mu,\nu,\cdots=0,1,2,3$ (Minkowskian metric) for the space-time
4-vectors; (iv) $\alpha,\beta,\cdots=0,1,2,3$ (Euclidean metric)
for isoscalar(0)+ isovectors(1,2,3).

\begin{table}
\caption{Parameters of the model Lagrangian}
\begin{center}
\begin{tabular}{ccc}
\hline
notation & physical meaning & value \\
\hline
$f_\pi$ & pion decay constant & 93 MeV \\
$f_\chi$ & $\chi$ decay constant & 210 MeV \\
$g$ & $\rho\pi\pi$ coupling constant &
5.85$^*$ \\
$m_\pi$  & pion mass & 140 MeV \\
$m_\chi$ & $\chi$ meass & 720 MeV \\
$m_V$ & vector meson masses &
770 MeV$^\dagger$  \\
$a$ & vector meson dominance & 2 \\
\hline
\multicolumn{3}{l}{\small $^*$ obtained
by using the KSFR relation $m_V^2=m_\rho^2=m_\omega^2=af_\pi^2
g^2$ with }\\
\multicolumn{3}{l}{\small \hskip 1em $a=2$.
cf. $g_{\rho\pi\pi}=6.11$ from the decay width of
$\rho\rightarrow\pi\pi$.} \\
\multicolumn{3}{l}{\small $^\dagger$ experimentally measured values
are $m_\rho$=768 MeV and $m_\omega$=782 MeV.}
\end{tabular}
\end{center}
\end{table}

Let us analyze the free space meson Lagrangian, i.e. the zero
baryon number ($B=0$) sector. The vacuum values of the fields are
given by,

\begin{equation}
U_{\mbox{\scriptsize vac}}=1, \hskip 2em
\rho^a_{\mu,\mbox{\scriptsize vac}}
 = \omega_{\mu,\mbox{\scriptsize vac}} = 0, \hskip 2em
\chi_{\mbox{\scriptsize vac}}=f_\chi.
\end{equation}
Letting the fields fluctuate in this vacuum through the Ansatz :
\begin{equation}
U \Leftarrow \exp(i \tau_a \tilde{\pi}_a), \hskip 2em
\rho^a_\mu \Leftarrow  \tilde{\rho}^a_\mu, \hskip 2em
\omega_\mu \Leftarrow \tilde{\omega}_\mu, \hskip 2em
\chi \Leftarrow f_\chi + \tilde{\chi}.
\label{Ans}\end{equation}
we create the mesons as can be seen by plugging the Ansatz
(\ref{Ans}) into the Lagrangian (\ref{lag}) and expanding in
fields and derivatives of the fields. We can then write the
Lagrangian as
\begin{equation}
{\cal L} =
   {\cal L}_{\mbox{\scriptsize free}} +
   {\cal L}_{\mbox{\scriptsize int}}
\label{Lfree}\end{equation}
where

$$ {\cal L}_{\mbox{\scriptsize free}} =
  \textstyle\frac12 \partial_\mu \tilde{\pi}^a \partial^\mu \tilde{\pi}^a
 - \frac12 m_\pi^2 \tilde{\pi}^a \tilde{\pi}^a
 -\textstyle\frac12 \partial_\mu \tilde{\rho_\nu}^a
\partial^\mu \tilde{\rho}^{a\nu}
 + \frac12 m_V^2 \tilde{\rho_\nu}^a \tilde{\rho}^{a,\nu} $$
$$\mbox{      }
- \textstyle\frac12 \partial_\mu \tilde{\omega}^\nu \partial^\mu
\tilde{\omega_\nu}
 + \frac12 m_V^2 \tilde{\omega_\nu} \tilde{\omega^\nu}
 +\textstyle\frac12 \partial_\mu \tilde{\chi} \partial^\mu \tilde{\chi}
 - \frac12 m_\chi^2 \tilde{\chi} \tilde{\chi}
\eqno(\mbox{\ref{Lfree}a})
$$
and

$${\cal L}_{\mbox{\scriptsize int}} =
g \varepsilon^{abc} \tilde{\rho}^a_\mu \tilde{\pi}^b
 \partial^\mu \tilde{\pi}^c + \cdots.
\eqno(\mbox{\ref{Lfree}b})$$
This Lagrangian provides physical meaning to the parameters of the
model as listed in Table 1. The term chosen to appear, as an
example, in Eq.(\ref{Lfree}b) determines the $\rho \rightarrow
\pi\pi$ decay width as
\begin{equation}
\Gamma_{\rho \rightarrow \pi\pi} = \frac23 \frac{g_{\rho\pi\pi}^2}{4\pi}
\frac{|q_m|^3}{m_\rho^2},
\end{equation}
where $|q_m|$ being the momentum of the pions in the decay rest
frame.

\section{The B=1 Skyrmion : Hedgehog Ansatz}

The solitons of these effective theories are the skyrmions. From
Eq.(\ref{lag}) the spherically symmetric hedgehog Ansatz for the
$B=1$ soliton solution of the standard Skyrme model can be
generalized to
\begin{equation}
U^{B=1} = \exp(i\vec{\tau}\cdot\hat{r} F(r)),
\label{UBeq1}\end{equation}
\begin{equation}
\rho^{a,B=1}_{\mu=i} = \varepsilon^{ika}\hat{r}^k \frac{G(r)}{gr},
\hskip 2em
\rho^{a,B=1}_{\mu=0} = 0,
\label{RBeq1}\end{equation}
\begin{equation}
\omega^{B=1}_{\mu=i} = 0,
\hskip 2em
\omega^{B=1}_{\mu=0} = f_\pi W(r),
\label{WBeq1}\end{equation}
and
\begin{equation}
\chi^{B=1} = f_\chi C(r).
 \label{CBeq1}\end{equation}
The boundary conditions that the profile functions satisfy at
infinity are
 \begin{equation}
F(\infty)=G(\infty)=W(\infty)=0, \hskip 2em
C(\infty)=1.
 \end{equation}

The Ans\"atze for $\rho$ and $\omega$ can be inferred from their
equations of motion by ignoring the space-time derivatives and
taking their masses to infinity. Thus for the $\rho$, from ${\cal
L}_{\pi\rho}$,  we get
 \begin{equation}
\rho^a_\mu \sim \frac{i}{g} \mbox{Tr}
\left[\frac{\tau^a}{2}(\ell_\mu + r_\mu)\right]
= \left\{
\begin{array}{ll}
\displaystyle \varepsilon^{iak} \hat{r}^k
\frac{1-\cos F(r)}{gr} & (\mu=i), \\
0 & (\mu=0),
\end{array} \right.
\label{Naive1}
 \end{equation}
and for the $\omega$, from  ${\cal L}_\omega + {\cal L}_{WZ}$, we
have
\begin{equation}
\omega_\mu \sim \frac{3g}{m_\omega^2} B_\mu
= \left\{
\begin{array}{ll}
0 & (\mu=i), \\
\displaystyle \frac{3g}{m_\omega^2} \frac{\sin^2 F(r)}{2\pi^2 r^2}
\frac{dF}{dr} & (\mu=0).
\end{array}\right.
\label{Naive2}\end{equation}
Equations (\ref{Naive1}-\ref{Naive2}), together with the behavior
of $F(r)$ near the origin,
\begin{equation}
F(r) = \pi - \alpha r - \textstyle \frac13 \gamma r^3 +\cdots,
\hskip 2em (r\ll 1)
\end{equation}
provide us with the  boundary conditions for $G(r)$ and $W(r)$ for
small $r$ :
\begin{equation}
G(0) = -2, \hskip 2em W^\prime(0)=0.
 \end{equation}

The equations of motion for the various profile functions $F(r)$,
$G(r)$, $W(r)$ and $C(r)$ can be gotten by minimizing the soliton
mass :
\begin{equation}
E_{B=1} = E_\pi^{B=1} + E_{\pi\rho}^{B=1} + E_\rho^{B=1}
        + E_\omega^{B=1} + E_{WZ}^{B=1} + E_\chi^{B=1},
\label{EBeq1}\end{equation}
where
$$E_\pi^{B=1} =
4\pi  \int_0^\infty r^2dr \frac{f_\pi^2}{2}
\left\{ C^2 \left( F^{\prime 2} + \frac{2}{r^2} \sin^2 F \right)
+ 2m_\pi^2 C^3(1-\cos F) \right\},
\eqno(\mbox{\ref{EBeq1}.a})$$
$$E_{\pi\rho}^{B=1} =
4\pi  \int^\infty_0 dr 2 f_\pi^2 C^2 (G+1-\cos F)^2,
\eqno(\mbox{\ref{EBeq1}.b})$$
$$E_\rho^{B=1} =
4\pi \int^\infty_0 dr \frac{1}{g^2}\left\{
G^{\prime 2} + \frac{G^2(G+2)^2}{2r^2} \right\},
\eqno(\mbox{\ref{EBeq1}.c})$$
$$E_\omega^{B=1} =
-4\pi \int_0^\infty r^2dr \frac{f_\pi^2}{2}
\left\{ W^{\prime 2} + m_\omega^2 C^2 W^2 \right\},
\eqno(\mbox{\ref{EBeq1}.d})$$
$$E_{WZ}^{B=1} =
4\pi \int_0^\infty r^2dr \frac{3g}{4\pi^2}f_\pi W
\frac{\sin^2 F}{r^2} F^\prime,
\eqno(\mbox{\ref{EBeq1}.e})$$
$$E_{\chi}^{B=1} =
4\pi \int_0^\infty r^2dr \frac{f_\chi^2}{2} \left\{
C^{\prime 2} + \textstyle \frac12 m_\chi^2(C^4(\ln C-\frac14)+\frac14)
\displaystyle \right\}.
\eqno(\mbox{\ref{EBeq1}.f})$$
They are,
\begin{eqnarray}
F^{\prime\prime} &=&
-\left(\frac{2}{r} + \frac{2C^\prime}{C} \right) F^\prime
+m_\pi^2 C \sin F
\nonumber\\
&&
+\frac{1}{r^2} \left( 4(G+1)\sin F - \sin 2F \right)
-\frac{3g}{4\pi^2 f_\pi} \frac{\sin^2 F}{r^2} W^\prime \frac{1}{C^2},
\label{Feq}\\
G^{\prime\prime} &=&
m_\rho^2 C^2 (G+1-\cos F) + \frac{G(G+1)(G+2)}{r^2},
\label{Geq}\\
W^{\prime\prime} &=&
-\frac{2}{r} W^\prime + m_\omega^2 C^2 W
- \frac{3g}{4\pi^2 f_\pi}\frac{\sin^2 F}{r^2} F^\prime,
\label{Weq}\\
C^{\prime\prime} &=&
-\frac{2}{r} C^\prime + m_\chi^2 C^3 \ln C \nonumber\\
&&+ \frac{f_\pi^2}{f_\chi^2}
\left\{ F^{\prime 2} + \frac{2}{r^2}\sin^2 F
    + 3 m_\pi^2 C (1-\cos F)
\right. \nonumber\\
&& \hskip 7em \left.
+\frac{4}{r^2} (G+1-\cos F)^2 - m_\omega^2 W^2 \right\} C.
\label{Ceq}
 \end{eqnarray}
Note that the contributions of $\omega$ to the mass,
$E_\omega^{B=1}$ and $E_{WZ}^{B=1}$, satisfy a virial
theorem \cite{zahedbrown}. Equation (\ref{EBeq1}.d) can be
expressed as
$$E_\omega^{B=1} =
-4\pi \int_0^\infty r^2dr \frac{f_\pi^2}{2} W \left\{ - W^{\prime
\prime}- \frac{2}{r} W^\prime + m_\omega^2 C^2 W \right\},$$
which implies that, if $W(r)$ satisfies the equation of motion
(\ref{Weq}),
 \be
-2E_\omega^{B=1}=E_{WZ}^{B=1}.
 \ee

The numerical results on the properties of the $B=1$ hedgehog
skyrmion (the mean square radius and the mass) are listed in Table
2 and the corresponding profile functions are given in Fig. 1. It
is interesting to see the roles the vector mesons and the dilaton
field play in describing the skyrmion. The $\omega$ meson provides
a strong repulsion and hence makes the soliton heavier and bigger
in size. Specifically, comparing the $\pi \rho$ model with the
$\pi \rho \omega$ model, we can see that when the presence of the
$\omega$ increases the mass by more than 415 MeV and the size,
i.e. $\la r^2\ra$, by more than .28 fm$^2$.

\begin{figure}[t]
\centerline{\epsfig{file=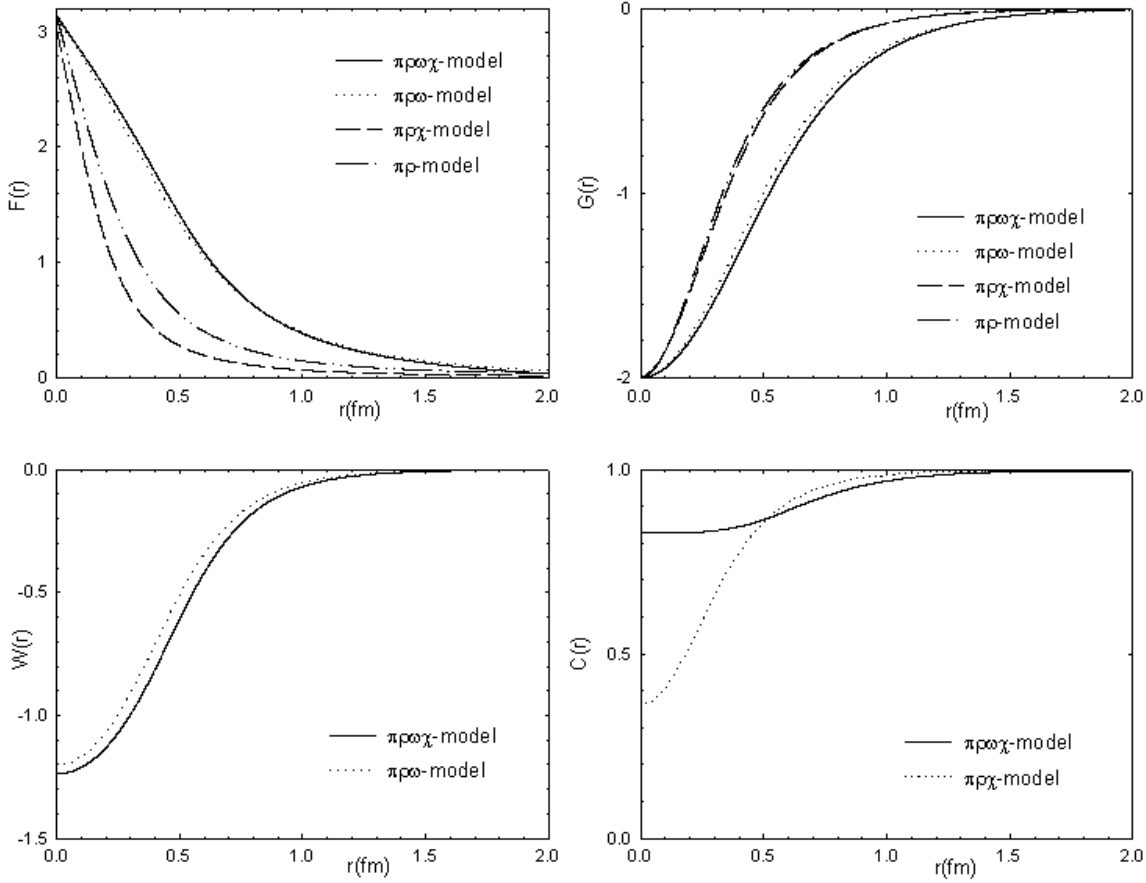,width=16cm,angle=0}}
\caption{Profile functions - $F(r)$, $G(r)$, $W(r)$ and $C(r)$.}
\end{figure}

How does the dilaton affect this calculation? The $\pi \rho$ model
with much smaller skyrmion has a larger baryon density near the
origin and this affects the dilaton, significantly changing its
mean-field value from its vacuum one.  The net effect of the
dilaton mean field on the mass is a reduction of $\sim$ 150 MeV,
whereas for the $\pi \rho \omega$ model it is only of 50 MeV. The
details can be seen in Table 2. The effect on the soliton size is,
however, different: while the dilaton in the $\pi \rho$ model
produces an additional localization of the baryon charge and hence
reduces $\la r^2\ra$ from .21 fm$^2$ to .19 fm$^2$, in the $\pi
\rho \omega$ model, on the contrary, the dilaton produces a
$delocalization$ and increases $\la r^2\ra$ from .49 fm$^2$ to .51
fm$^2$.

In sum, we note that when the $\omega$ is present, it plays a
major role in the skyrmion properties with the dilaton field
playing a minor role.

\begin{table}
\caption{Single skyrmion mass and various contributions to it.}
\begin{center}
\begin{tabular}{ccccccccc}
\hline
Model & $\langle r^2\rangle$ & $E^{B=1}$ & $E^{B=1}_\pi$ & $E^{B=1}_{\pi\rho}$
& $E^{B=1}_\rho$ & $E^{B=1}_\omega$ & $E^{B=1}_{WZ}$ & $E^{B=1}_\chi$ \\
\hline
$\pi\rho$-model & 0.27 & 1054.6 & 400.2 + 9.2 & 110.4 & 534.9
& 0.0 & 0.0 & 0.0 \\
$\pi\rho\chi$-model & 0.19 & 906.5 & 103.1 + 1.4 & 155.1 & 504.1
& 0.0 & 0.0 & 142.8 \\
$\pi\rho\omega$-model & 0.49 & 1469.0 & 767.6 + 39.9 & 33.2 & 370.7
  & -257.6 & 515.1 & 0.0 \\
$\pi\rho\omega\chi$-model & 0.51 & 1408.3 & 646.0 + 29.2 & 34.9 & 355.7
  & -278.3 & 556.7 & 64.2 \\
\hline
\end{tabular}
\end{center}
\end{table}

\section{The B=2 Skyrmion : Product Ansatz}

In the original Skyrme model with only pion fields, we can obtain
a $B=2$ configuration by simply taking the product of two $B=1$
hedgehog $U$ fields. This configuration can be a good
approximation to the true solution when the two skyrmions are
sufficiently far apart. Furthermore the energy of this
configuration becomes the lowest when one of the skyrmions is
rotated in isospin space by an angle $\pi$ with respect to the
axis perpendicular to the line joining two skyrmions. Here we have
to generalize this feature to incorporate the $\rho$, $\omega$ and
$\chi$ fields in the scheme. When two skyrmions are sufficiently
far apart, our Ansatz should describe each individual skyrmion
correctly. Recalling the field configurations for a single
skyrmion at infinity, the most natural Ansatz is
\begin{equation}
\begin{array}{c}
\displaystyle U_{B=2} \Leftarrow
U_{B=1}(\vec{r}_1)*U_{B=1}(\vec{r}_2), \\
\displaystyle
\rho^{a,B=2}_\mu \Leftarrow \rho^{a,B=1}_\mu(\vec{r}_1)
                  + \rho^{a,B=1}_\mu(\vec{r}_2), \\
\displaystyle
\omega^{B=2}_\mu \Leftarrow \omega^{B=1}_\mu(\vec{r}_1)
                 + \omega^{B=1}_\mu(\vec{r}_2), \\
\displaystyle
\chi^{B=2} \Leftarrow \chi^{B=1}(\vec{r}_1)*\chi^{B=1}(\vec{r}_2),
\end{array}
\label{Simple1}\end{equation}
where $\vec{r}_{1,2}$ stand for the position of the centers of
each skyrmion. The boundary conditions for $B=1$ skyrmion imply
that a {\em multiplicative} rule must apply to $U$ and $\chi$,
since their vacuum values at infinity are 1, and an {\em additive}
rule to $\rho$ and $\omega$, since theirs vanish at infinity.

The relative orientation takes place in isospin space. Thus the
isoscalar fields, $\omega$ and $\chi$, do not undergo rotation.
However the isovector  fields $\pi$ and $\rho$ could require
non-trivial relative orientations.
\begin{equation}
\begin{array}{c}
\displaystyle
U_{B=2} = U_{B=1}(\vec{r}_1)* AU_{B=1}(\vec{r}_2)A^\dagger, \\
\\
\displaystyle
\vec{\tau}\cdot\vec{\rho}^{B=2}_\mu
= \vec{\tau}\cdot\vec{\rho}^{B=1}_\mu(\vec{r}_1)
+ A\vec{\tau}\cdot\vec{\rho}^{a,B=1}_\mu(\vec{r}_2)A^\dagger,
\end{array}
\label{Simple2}\end{equation}
where $A$ is an SU(2) matrix. This  Ansatz is a drastically
simplified one and will not produce the lowest energy
configuration of the $B=2$ system, known to be of the toroidal
shape \cite{manton}. However for the purpose of our calculation,
we need only the Ansatz defined by Eqs. (\ref{Simple1}) and
(\ref{Simple2}), which will be used  to define the starting
configuration of the skyrmion matter \cite{skyrmionmatter}. As in
the original Skyrme model, the lowest energy for this type of
Ansatz can be obtained in the configuration where two skyrmions
are relatively rotated in isospin space by an angle $\pi$ with
respect to an axis perpendicular to the line joining them. (See
Fig.2.)

In Fig.3, we show the density profile of the lowest energy
configuration, in this approximation, which is of dumbbell shape
with well-separated skyrmions.

\begin{figure}[tbp]
\centerline{\epsfig{file=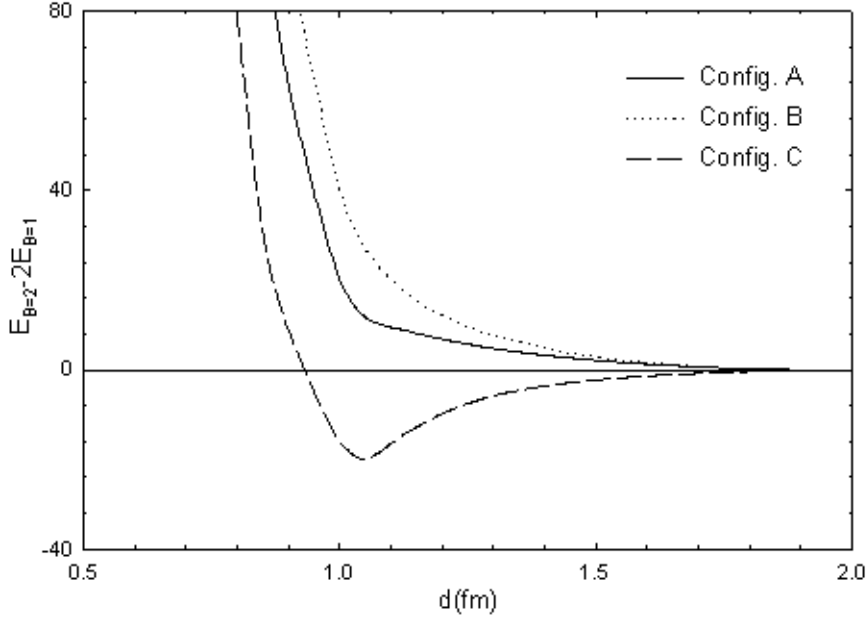,width=12cm,angle=0}}
\caption{Binding energy of the two-skyrmion system. Configuration
A represents the case with no relative rotation. Configuration B
represents the case where the skyrmion is rotated by an angle
$\pi/2$ with respect to the axis parallel to the line joining the
two skyrmions. In configuration  C,  the angle is $\pi$ and the
rotation axis is perpendicular to the line joining the two
skyrmions.}
\end{figure}

\begin{figure}
\centerline{\epsfig{file=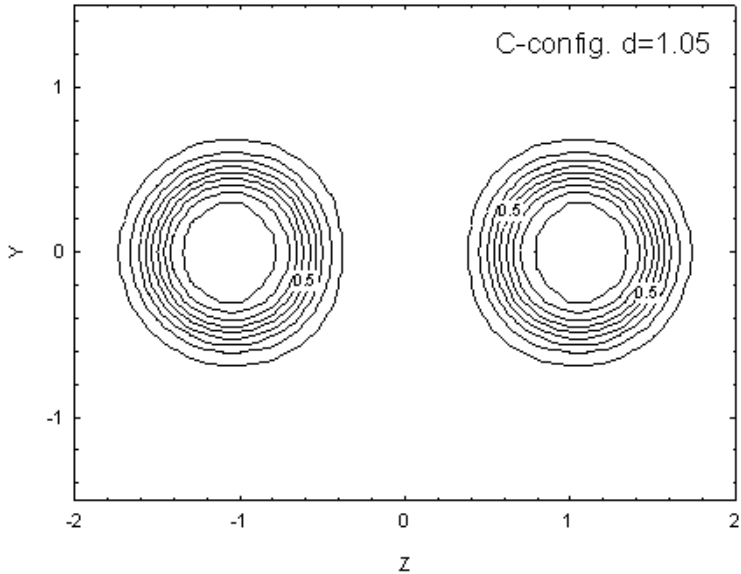,width=12cm,angle=0}}
\caption{The lowest energy configuration obtained under the naive
product Ansatz (\ref{Simple1}).}
\end{figure}

\section{Skyrmion Matter : an FCC skyrmion crystal}

We have learned from the study on two-skyrmion system in Sec.4
that the lowest-energy configuration is obtained when one of the
skyrmions is rotated in isospin space with respect to the other by
an angle $\pi$ about an axis perpendicular to the line joining the
two. If we generalize this Ansatz to many-skyrmion matter, we
obtain that the configuration at the classical level for a given
baryon number density is an FCC crystal where the nearest neighbor
skyrmions are arranged to have the attractive relative
orientations \cite{skyrmionmatter}~\footnote{An early discussion
on the FCC crystal structure of nuclei can be found in
\cite{FCC_in_NP}. See also \cite{garai} for a recent discussion. A
quark-model description -- which is similar in spirit to our
approach -- in which the gluonic mean field represented by a
confining potential replaces the skyrmion field distribution is
given by Goldman et al~\cite{FCC_in_QP} and Kim Maltman et
al~\cite{kimmaltman}. We are grateful to Terry Goldman for
bringing our attention to these references.}. We next proceed to
incorporate the vector mesons by generalizing Kugler's Fourier
series expansion method \cite{KS89}, developed for the original
Skyrme model, to our $\pi\rho\omega\chi$-model. The symmetries of
such FCC crystal configuration are summarized in Table 3. As for
the $\rho$-fields, we find it more convenient to give the
symmetries for the dual vector fields $\tilde{\rho}_p$ as defined
in Table 3. Note: (i) $\tilde{\rho}_a(a=1,2,3)$ has the same
symmetries as $\pi_a$; (ii) the isoscalar fields $\sigma$,
$\omega_0$ and $\chi$ share the same symmetries. One can easily
check the symmetries for $\rho^a_i$(or $\tilde{\rho}_a$) and
$\omega_0$ by referring to the useful relations (\ref{Naive1}) and
(\ref{Naive2}).

\begin{table}[tbp]
\begin{center}
\caption{Symmetries of the FCC skyrmion crystal}
\begin{tabular}{cccccc}
\hline
&
& $\begin{array}{c}\mbox{reflection}\\ \mbox{($yz$-plane)}\end{array}$
& $\begin{array}{c}\mbox{3-fold axis}\\ \mbox{rotation}\end{array}$
& $\begin{array}{c}\mbox{4-fold axis}\\ \mbox{($z$-axis) rot.}\end{array}$
& translation \\
\hline
$(x,y,z)$
& $\rightarrow$
& $(-x,y,z)$
& $(y,z,x)$
& $(x,z,-y)$
& $(x+L,y+L,z)$ \\
$U=\sigma+i\vec{\tau}\cdot\vec{\pi}$
& $\rightarrow$
& $(\sigma, -\pi_1, \pi_2, \pi_3)$
& $(\sigma, \pi_2, \pi_3, \pi_1)$
& $(\sigma, \pi_1, \pi_3, -\pi_2)$
& $(\sigma, -\pi_1, -\pi_2, \pi_3)$ \\
$\rho^a_i \equiv \varepsilon_{aip} \tilde{\rho}_p $
& $\rightarrow$
& $(-\tilde{\rho}_1,\tilde{\rho}_2,\tilde{\rho}_3)$
& $(\tilde{\rho}_2,\tilde{\rho}_3,\tilde{\rho}_1)$
& $(\tilde{\rho}_1,\tilde{\rho}_3,-\tilde{\rho}_2)$
& $(-\tilde{\rho}_1,-\tilde{\rho}_2,\tilde{\rho}_3)$ \\
$\omega_0,\chi$
& $\rightarrow $
& $\omega_0,\chi$
& $\omega_0,\chi$
& $\omega_0,\chi$
& $\omega_0,\chi$ \\
\hline
\end{tabular}
\end{center}
\end{table}

\subsection{Fourier series expansion}

We next generalize our work in
ref.(\cite{skyrmionmatter,sliding}), following Kugler and
Shtrikman\cite{KS89}, to the problem at hand by indicating the
pertinent expansions for all the fields.
 \begin{itemize}
\item[i)] {\bf The pion expansion}:

We obtain the pion fields\footnote{We use
$\phi^\pi_\alpha(\alpha=0,1,2,3)$ instead of
$\sigma(=\phi^\pi_0)$, $\pi_a(=\phi^\pi_a,{}a=1,2,3)$. The
superscript $\pi$ denotes that these fields are associated with
the pion fields. We will introduce similar fields $\phi_\mu^\rho$
for the $\rho$ fields.}  $\phi^\pi_\alpha $ from the un-normalized
fields $\bar{\phi}^\pi_\alpha$ which are expanded in Fourier
series as

\begin{equation}
\begin{array}{rcl}
\bar{\phi}^\pi_0 &=&
\displaystyle
\sum_{abc} \beta^\pi_{abc} \cos(\pi ax/L)\cos(\pi by/L)\cos(\pi cz/L), \\
\bar{\phi}^\pi_1 &=&
\displaystyle
\sum_{hkl} \alpha^\pi_{hkl} \sin(\pi hx/L)\cos(\pi ky/L)\cos(\pi lz/L), \\
\bar{\phi}^\pi_2 &=&
\displaystyle
\sum_{hkl} \alpha^\pi_{hkl} \cos(\pi lx/L)\sin(\pi hy/L)\cos(\pi kz/L), \\
\bar{\phi}^\pi_3 &=&
\displaystyle
\sum_{hkl} \alpha^\pi_{hkl} \cos(\pi kx/L)\cos(\pi ly/L)\sin(\pi hz/L),
\end{array}
\label{pi}\end{equation}
and thereafter normalized

\begin{equation}
\phi^\pi_\alpha = \frac{\bar{\phi}^\pi_\alpha}
{\sqrt{\sum_\beta (\bar{\phi}^\pi_\beta)^2}}.
\end{equation}
In order to be consistent with the symmetry properties listed in
Table 3, $(a,b,c)$ should be all even or all odd integers, and
$(k,l)$ should be all even(odd) if $h$ is odd(even). Furthermore,
to provide a correct topological structure to the configuration,
the expansion coefficients $\beta_{abc}$ must satisfy the
constraint,
\begin{equation}
\sum_{\mbox{\scriptsize even}}\beta^\pi_{abc} = 0.
\end{equation}

\item [ii)] {\bf The $\rho$  expansion} :

We introduce $\phi^\rho_\alpha (\alpha=0,1,2,3)$ and using
Eq.(\ref{Naive1}), we may express $\rho^a_i$ in terms of these
fields
 \begin{equation}
\rho^a_i = \varepsilon_{abc} \hat{\phi}^\rho_b \partial_i
\hat{\phi}^\rho_c (1-\phi^\rho_0)
= \varepsilon_{abc} {\phi}^\rho_b \partial_i
{\phi}^\rho_c /(1+\phi^\rho_0) ,
\end{equation}
where $\hat{\phi}^\rho_a=\phi^\rho_a/\sqrt{\sum_{b=1}^3
(\phi^\rho_b)^2} =\phi^\rho_a/\sqrt{1-(\phi^\rho_0)^2}$. Similarly
to the pion fields, $\phi^\rho_\alpha$ is obtained from the
un-normalized fields $\bar{\phi}^\rho_\alpha$ whose expansion in
Fourier series is
\begin{equation}
\begin{array}{rcl}
\bar{\phi}^\rho_0 &=&
\displaystyle
\sum_{abc} \beta^\rho_{abc} \cos(\pi ax/L)\cos(\pi by/L)\cos(\pi cz/L), \\
\bar{\phi}^\rho_1 &=&
\displaystyle
\sum_{hkl} \alpha^\rho_{hkl} \sin(\pi hx/L)\cos(\pi ky/L)\cos(\pi lz/L), \\
\bar{\phi}^\rho_2 &=&
\displaystyle
\sum_{hkl} \alpha^\rho_{hkl} \cos(\pi lx/L)\sin(\pi hy/L)\cos(\pi kz/L), \\
\bar{\phi}^\rho_3 &=&
\displaystyle
\sum_{hkl} \alpha^\rho_{hkl} \cos(\pi kx/L)\cos(\pi ly/L)\sin(\pi hz/L),
\end{array}
\end{equation}
which we then normalize. The procedure is analogous to that of the
pion fields, although the expansion coefficients are different.

\item [iii)] {\bf The $\omega$ and $\chi$ expansions} :

The isoscalar fields $\omega_0$ and $\chi$ have the same symmetry
properties as the $\sigma$ and therefore their expansions are,
\begin{equation}
\begin{array}{rcl}
\omega_0/f_\pi \equiv w&=&
\displaystyle
\sum_{abc} \beta^\omega_{abc} \cos(\pi ax/L)\cos(\pi by/L)\cos(\pi cz/L), \\
\chi/f_\chi \equiv c&=&
\displaystyle
\sum_{abc} \beta^\chi_{abc} \cos(\pi ax/L)\cos(\pi by/L)\cos(\pi cz/L). \\
\end{array}
\label{w_and_c}\end{equation}
\end{itemize}
The expansion coefficients are determined such that the energy per
skyrmion is minimized:
\begin{equation}
E/B = (E/B)_\pi + (E/B)_{\pi\rho} + (E/B)_\rho
    + (E/B)_\omega + (E/B)_{WZ} + (E/B)_\chi.
\label{EperB}\end{equation}
where
$$(E/B)_\pi =
\textstyle \frac14 \displaystyle \int_{\mbox{\scriptsize Box}} d^3x
\frac{f_\pi^2}{2} c^2  \left\{ (\partial_i \phi^\pi_\alpha)^2
+2m_\pi^2 c (1-\phi^\pi_0)\right\},
\eqno(\mbox{\ref{EperB}.a})$$
$$(E/B)_{\pi\rho} =
\textstyle \frac14 \displaystyle \int_{\mbox{\scriptsize Box}} d^3x
2f_\pi^2 c^2 \left\{
\frac{\vec{\phi}^\pi \times \partial_i \vec{\phi}^\pi}
{1+\phi^\pi_0}
-\frac{\vec{\phi}^\rho \times \partial_i \vec{\phi}^\rho}
{1+\phi^\rho_0} \right\}^2,
\eqno(\mbox{\ref{EperB}.b})$$
$$(E/B)_{\rho} =
\textstyle \frac14 \displaystyle \int_{\mbox{\scriptsize Box}} d^3x
\frac{1}{8g^2} \left(
\partial_i \phi^\rho_\alpha \partial_j \phi^\rho_\beta
-\partial_j \phi^\rho_\alpha \partial_i \phi^\rho_\beta
\right)^2,
\eqno(\mbox{\ref{EperB}.c})$$
$$(E/B)_{\omega} =
-\textstyle \frac14 \displaystyle \int_{\mbox{\scriptsize Box}} d^3x
\textstyle \frac12 \left\{
(\partial_i w_0)^2 + m_\omega^2 c^2 w_0^2 \right\},
\eqno(\mbox{\ref{EperB}.d})$$
$$(E/B)_{WZ} =
\textstyle \frac14 \displaystyle \int_{\mbox{\scriptsize Box}} d^3x
\frac{3g}{2}f_\pi w_0 B_0,
\eqno(\mbox{\ref{EperB}.e})$$
$$(E/B)_{\chi} =
\textstyle \frac14 \displaystyle \int_{\mbox{\scriptsize Box}} d^3x
\frac{f_\chi^2}{2} \left\{
(\partial_i c)^2
+ \textstyle\frac12 m_\chi^2 [ c^4(\ln c-\frac14)+\frac14]
\right\},
\eqno(\mbox{\ref{EperB}.f})$$
and
$$B_0 = \frac{1}{12\pi^2}\varepsilon_{ijk}
\varepsilon_{\alpha\beta\gamma\delta}
\phi^\pi_\alpha \partial_i \phi^\pi_\beta
\partial_j \phi^\pi_\gamma \partial_k \phi^\pi_\delta
=
\frac{1}{2\pi^2} \left|
\begin{array}{cccc}
\phi^\pi_0 & \phi^\pi_1 & \phi^\pi_2 & \phi^\pi_3 \\
\partial_1 \phi^\pi_0 & \partial_1 \phi^\pi_1
  & \partial_1 \phi^\pi_2 & \partial_1 \phi^\pi_3 \\
\partial_2 \phi^\pi_0 & \partial_2 \phi^\pi_1
  & \partial_2 \phi^\pi_2 & \partial_2 \phi^\pi_3 \\
\partial_3 \phi^\pi_0 & \partial_3 \phi^\pi_1
  & \partial_3 \phi^\pi_2 & \partial_3 \phi^\pi_3
\end{array} \right|.
\eqno(\mbox{\ref{EperB}.g})$$

Since there are no constraints on the expansion parameters for the
$\omega$ and $\chi$, a straightforward variational process of
minimizing the energy fails. Note that as far as the energy of the
system (\ref{EperB}) is concerned, $\omega_0=0$ comes out as the
solution which minimizes the energy \footnote{This solution cannot
of course be the true solution of the model as it does not satisfy
the equations of motion eq.(\ref{w_eq}) containing the
inhomogeneous source term.}. This is confirmed in Table 1 where we
see that without the repulsive $\omega$, the single skyrmion mass
becomes much lower. Therefore, the variational process always
leads us to $w_0=0$, which corresponds to the energy per baryon of
the $\pi \rho \chi$ model and not of the present model.

We realize that the $\omega$ needs a careful treatment in the
model that incorporates it self-consistently. We proceed therefore
to treat the $\omega$ in a more elaborate way.  Instead of
including the expansion coefficients $\beta^\omega_{abc}$ into the
minimization process, we fix them by solving the equation of
motion for $w (=w_0/f_\pi)$ at each step. The equation of motion
derived from (\ref{EperB}) becomes
\begin{equation}
(-\partial_i^2 + m_\omega^2 c^2 ) w = -\frac{3g}{2f^2_\pi} B_0.
\label{w_eq}\end{equation}
Note that $c^2$ and $B_0$ have the same symmetry structure.
Explicitly, the expansions (\ref{pi}) and (\ref{w_and_c}) lead to
\begin{equation}
c^2=
\sum_{abc} \beta^{c^2}_{abc} \cos(\pi ax/L)\cos(\pi by/L)\cos(\pi cz/L),
\end{equation}
\begin{equation}
B_0=
\sum_{abc} \beta^{B_0}_{abc} \cos(\pi ax/L)\cos(\pi by/L)\cos(\pi cz/L),
\end{equation}
with the expansion coefficients $\beta^{c^2}_{abc}$ and
$\beta^{B_0}_{abc}$ determined from the configuration of
$\phi^\pi_\alpha$ and $c$. The equation of motion, therefore,
reduces to a linear equation for the expansion coefficients
$\beta^{\omega}_{abc}$ given by
\begin{equation}
\sum_{a^\prime b^\prime c^\prime} M_{abc,a^\prime b^\prime c^\prime}
\beta^\omega_{a^\prime b^\prime c^\prime}
=\frac{3g}{2f_\pi^2} \beta^{B_0}_{abc}.
\end{equation}
The matrix elements are
\begin{equation}
M_{abc,a^\prime b^\prime c^\prime}
=(a^2+b^2+c^2)\left(\frac{\pi}{L}\right)^2 \delta_{abc,a^\prime
b^\prime c^\prime} +m^2_\omega \sum_{a^{\prime\prime}
b^{\prime\prime} c^{\prime\prime}} \beta ^{c^2}_{a^{\prime\prime}
b^{\prime\prime} c^{\prime\prime}} f_{a^\prime,
a^{\prime\prime},a} f_{b^\prime, b^{\prime\prime},b} f_{c^\prime,
c^{\prime\prime},c},
\end{equation}
where

$$ f_{a^\prime, a^{\prime\prime},a} = \left\{
\begin{array}{ll}
\delta_{a^\prime,a}, & \mbox{if $a^{\prime\prime}=0$}, \\
\delta_{a^{\prime\prime},a}, & \mbox{if $a^\prime=0$}, \\
\frac12 \delta_{a,a^\prime \pm a^{\prime\prime}},
  & \mbox{if $a^\prime \neq 0$ and $a^{\prime\prime}\neq 0$}.
\end{array}
\right.
$$

As formulated, our variational procedure is restricted to evolve
in the space of parameters which satisfy the $\omega$ equations of
motion and therefore the ``false" solution is not present in the
scheme.

\section{Results}
We show in Figs. 4 and 5 the the numerical results of the energy
per baryon $E/B$, $\langle \chi\rangle$ and $\langle
\sigma\rangle$ in various models as a function of the FCC lattice
parameter $L$ \footnote{The baryon number density is given by
$\rho_B = 1/2L^3$. Normal nuclear matter density $\rho_0 =
0.17/\mbox{fm}^3$ corresponds to $L\sim 1.43$.}.

In the $\pi \rho \chi$ model, as the density of the system
increases ($L$ decreases), $E/B$ changes little (see Fig. 4). It
is close to the energy of a $B=1$ skyrmion up to a density greater
than $\rho_0 \; (L \sim 1.43)$. This result is easy to interpret.
As we discussed in Sec. 3, the size of the skyrmion in this model
is very small and therefore the skyrmions in the lattice will
interact only at very high densities, high enough for their tails
to overlap.

In the absence of the $\omega$, the dilaton field plays a dramatic
role. A skyrmion matter undergoes an abrupt phase transition at
high density at which the expectation value of the dilaton field
vanishes $\langle \chi \rangle=0$ \footnote{In general, $\langle
\chi\rangle =0$ does not necessarily require $\langle \chi^2
\rangle =0$. However, in our numerical results, $\langle
\chi\rangle =0$ always accompanies $\chi=0$ in the whole space.}.

In the  $\pi \rho \omega \chi$ model, the situation changes
dramatically as can be seen in Fig.5. The reason is that the
$\omega$ provides not only a strong repulsion among the skyrmions,
but somewhat surprisingly, also an intermediate range attraction.
Note the different mass scales between Figs. 4 and 5. In both the
$\pi\rho\omega$ and the $\pi\rho\omega\chi$ models, at high
density, the interaction reduces $E/B$ to 85\% of the $B=1$
skyrmion mass. This value should be compared with 94\% in the
$\pi\rho$ model. In the $\pi\rho\chi$-model, $E/B$ goes down to
74\% of the $B=1$ skyrmion mass, but in this case it is due to the
dramatic behavior of the dilaton field.

In the $\pi\rho\omega\chi$ model the role of the dilaton field is
suppressed. It provides a only a small attraction at intermediate
densities. Moreover, the phase transition towards its vanishing
expectation value, $\langle \chi\rangle=0$, does not take place.
Instead, its value grows at high density!

In all the models analyzed thus far, $\langle \sigma \rangle$ goes
to zero at high density. Thus, we have two different scenarios for
the chiral transition. The models without the $\omega$, but with
$\chi$, tend to bring about a phase transition that is consistent
with the ``vector manifestation" scenario, with the pion decay
constant $f_\pi$ vanishing. However, once the $\omega$ is present,
the $f_\pi$ remains non-zero while $\la \sigma\ra$ vanishes, a
scenario that is reminiscent of the pseudo-gap realization. This
is at variance with the VM scenario.

\begin{figure}[tbp]
\centerline{\epsfig{file=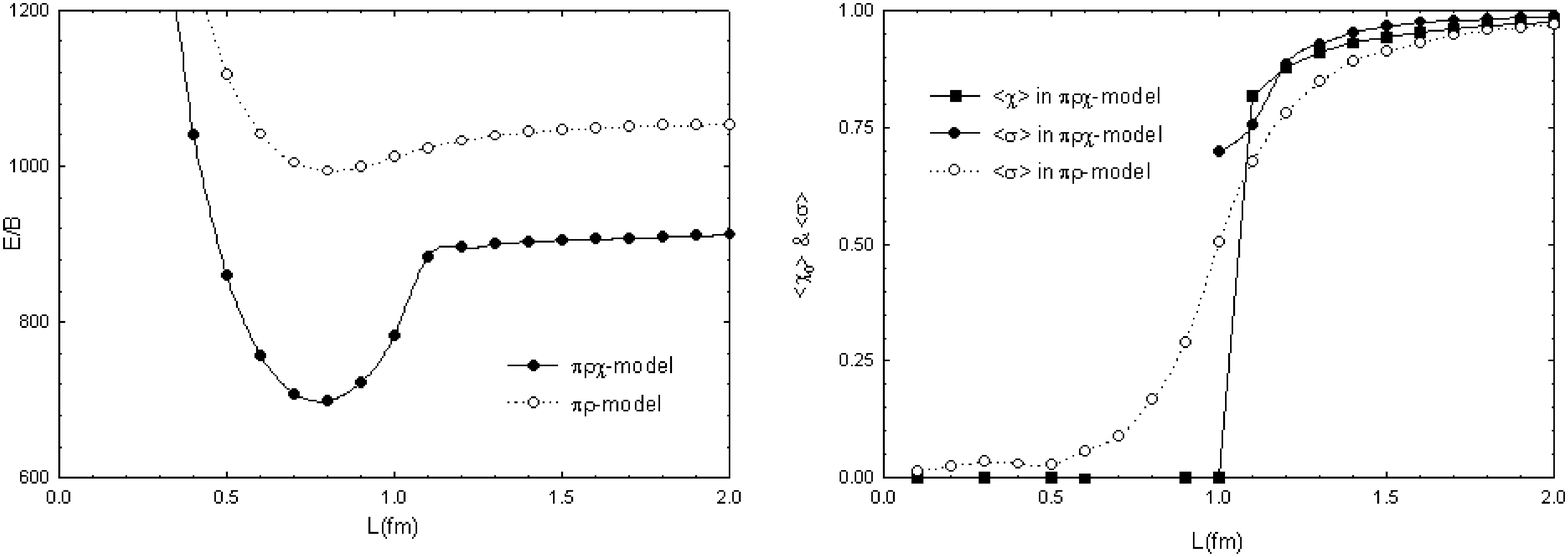,width=16cm,angle=0}}
 \caption{$E/B$,
$\langle \chi\rangle$ and $\langle\sigma\rangle$ as a function of
$L$ in the models without the $\omega$.}
\end{figure}
\begin{figure}
\centerline{\epsfig{file=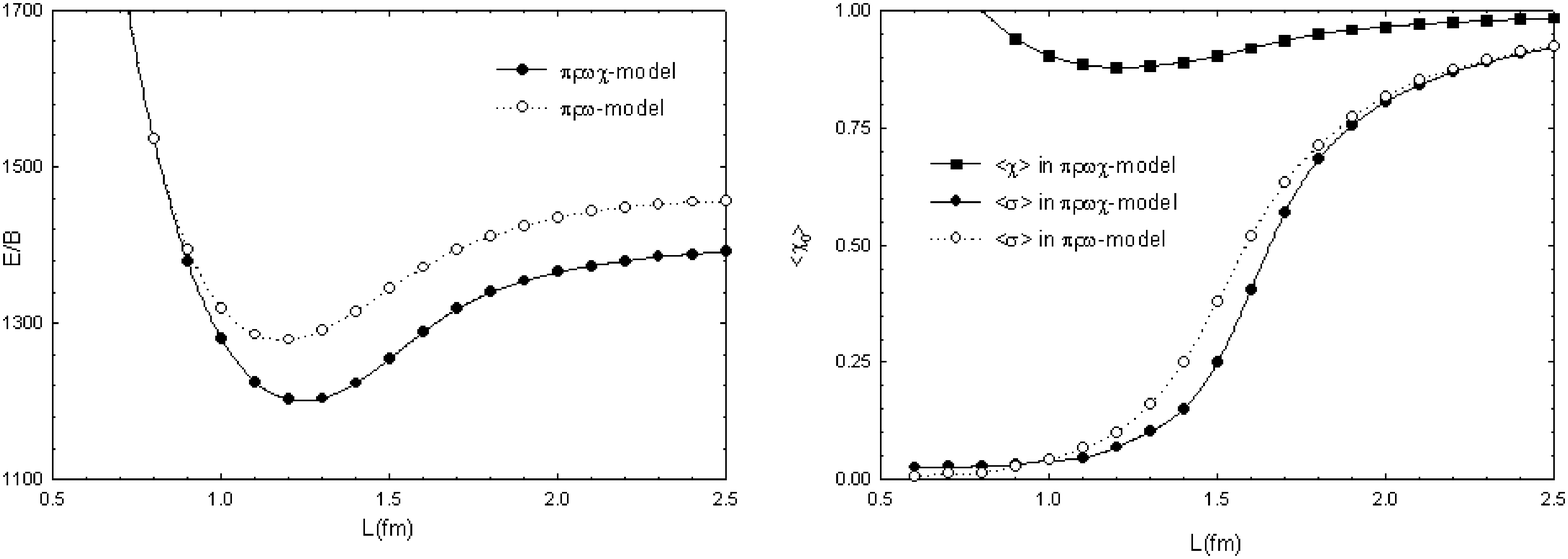,width=16cm,angle=0}}
\caption{$E/B$,
$\langle \chi\rangle$ and $\langle\sigma\rangle$ as a function of
$L$ in the models with the $\omega$.}
\end{figure}

Let us discuss in detail why and how the $\omega$ changes the
VM-like behavior of the $\pi \rho \chi$ model at the phase
transition into a pseudo-gap scenario. The term in the Lagrangian
responsible for this behavior is the coupling of the $\omega$ to
the topological current $B_0$, i.e. Eqs.(\ref{EperB}.e) and
(\ref{EperB}.g). To see this point, we write the equation of
motion for the $\omega$ as

$$
(-\partial_i^2 + m_\omega^{*2}) w = -\frac{3g}{2f_\pi^2} B_0,
$$
where we have replaced the space-dependent mass term $m_\omega^2
c^2$ of Eq.(\ref{w_eq}) by an effective mass $m_\omega^{*2}$. This
equation can be exactly solved by means of the appropriate Green's
function,
\begin{equation}
w = -\frac{3g}{2f_\pi^2}\int d^3 x^\prime \frac{\exp(-m_\omega^*
|\vec{x}-\vec{x}^\prime|)} {4\pi|\vec{x}-\vec{x}^\prime|}
B_0(\vec{r^\prime}). \label{omegaeqofmotion}
\end{equation}
Thus, $w$ can be interpreted as a static potential generated by
the source $B_0$. The $\omega$ contribution to $E/B$ can be
expressed as
\begin{equation}
(E/B)_{WZ} = \textstyle \frac14 \displaystyle
\int_{\mbox{\scriptsize Box}} d^3x \int d^3 x^\prime
\left(\frac{3g}{2}\right)^2 B_0(\vec{x}) \frac{\exp(-m_\omega^*
|\vec{x}-\vec{x}^\prime|)} {4\pi|\vec{x}-\vec{x}^\prime|}
B_0(\vec{x}^\prime), \label{green}
\end{equation}
and the term $(E/B)_\omega$ is, as mentioned previously, $-
\frac{1}{2}$ of this. Note that while the integral over $\vec{x}$
is defined in a single FCC cell, that over $\vec{x}^\prime$ is
not. Thus, unless it is screened, the periodic source $B_0$
filling infinite space will produce an infinite potential $w$
which leads to an infinite $(E/B)_{WZ}$. The screening is done by
the omega mass, $m_\omega^*$, as can be seen from
Eq.(\ref{green}). Thus the effective $\omega$ mass cannot vanish
for the solution! Our numerical results reflect this fact: at high
density the $B_0$-$B_0$ interaction becomes large compared to any
other contribution. In order to reduce it, $\chi$ has to increase,
and thereby the effective screening mass $m_\omega^* \sim m_\omega
\langle\chi\rangle$ becomes larger. In this way we run into a
phase transition where the expectation value of $\chi$ does not
vanish and therefore $f_\pi$ does not vanish but instead
increases.

\section{Concluding remarks}

This paper represents the natural continuation of our effort to
understand the physics of nuclear matter from the Skyrme model
description.  The (original) Skyrme model is an effective theory
which represents the large $N_c$ limit of $QCD$. In some sense, it
is the lowest order of a bosonized approach to this theory. Higher
orders will be obtained by incorporating massive mesons. In
particular, we know from phenomenology that the vector mesons play
a crucial role in describing the N-N interaction
\cite{brownjackson} and infinite nuclear matter~\cite{walecka}.
Thus going beyond our previous developments
\cite{skyrmionmatter,sliding}, the next natural step is to
implement vector mesons in consistency with $QCD$
\cite{weinberg,witten}. The model we studied in this paper
contains, in addition to the Goldstone pions, the scale dilaton
associated with trace anomaly of QCD and the $\rho$ and $\omega$
fields introduced according to the hidden local symmetry strategy.

As in our previous work without vector mesons, the skyrmion matter
possesses two phases: a low-density phase which we simply describe
here by an FCC crystal \footnote{We refer the reader to our
previous work for the discussion related to how we treat this
unstable phase.} and a high-density phase which is described by a
half-skyrmion CC crystal. In our previous work, the dilaton was
crucial to realize the phase transition in a scenario consistent
with the vector manifestation (VM) fixed point structure: The
phase transition was signalled by the vanishing of both
$\la\sigma\ra$ and $f_\pi^*$. However in the present model with
the vector mesons, in particular with the $\omega$, the mean
$\chi$ field cannot vanish and therefore, although $\la\sigma\ra$
vanishes, $f_\pi^*$ does not and we seem to fall back to a
pseudo-gap-type of picture which does not seem consistent with the
VM and in that matter with the standard sigma-model scenario. The
sole agent responsible for this puzzling feature is the $\omega$
meson.

If the $\omega$ is not present, i.e., in the $\pi \rho \chi$
model, the behavior is the conventional one. The $\chi$ field
vanishes at the phase transition and chiral restoration is
produced in the standard scenario. The $\rho$ meson is basically a
spectator at the classical level, producing little change with
respect to our previously studied $\pi \chi$ model
\cite{sliding,pionvel} except that at high densities, once the
$\rho$ starts to overlap, the energy of nuclear matter increases
due to its the repulsive effect at short distances. The densities
have to be quite high since these skyrmions are very small. Since
$\chi$ vanishes at the phase transition, we recover the VM
behavior, namely, $f_\pi^*=0$  and $m_\rho^*=0$.

The incorporation of the $\omega$ changes dramatically the scales.
Since the $\omega$ produces a strong repulsion, the skyrmions
become more massive and much bigger in size \cite{zahedbrown}.
Moreover the phase transition scenario changes dramatically from
the previously described. In particular, $\chi$ does not vanish at
the phase transition but it even increases its value. The
mechanism of how this happens is simple and robust. It is that the
$\omega$ coupling to the baryon density in our scheme leads in
nuclear matter to a long range infinite interaction unless some
sort of screening intervenes. However in medium the effective
$\omega$ mass becomes $m_\omega^* = m_\omega \chi$. Thus if $\chi$
decreases, the effective mass decreases and the screening
decreases, thus the long-range interaction becomes stronger and
ultimately will tend to dominate. In order to prevent this from
happening, the background skyrmion counteracts so as to compensate
for the increase in the baryon density interaction: It increases
the value of the $\chi$ field and hence the effective $\omega$
mass, i.e., the screening. Since the in-medium value of the pion
decay constant is locked to the mean-field value of $\chi$ which
does not vanish at the chiral transition,  $f_\pi$ does not
vanish, whereas the $\la\sigma\ra$, whose vanishing is linked to
the structure of the crystal, does to vanish at the critical
point, signalling the phase transition as in our naive scenario
\cite{skyrmionmatter}.

The qualitatively dramatic effect of the $\omega$ meson in the
skyrmion structure of dense matter is at the same time disturbing
and puzzling. In nuclear physics, $\omega$ has been a crucial
element~\cite{walecka}, so it makes one wonder what goes wrong at
densities greater than that of normal nuclear matter. We have no
clear answer at this point but we can think of three
possibilities: (a) the minimal Lagrangian we use is inadequate in
that short-distance physics is not $fully$ accounted for. Since
the $\omega$ degree of freedom accounts for nuclear interactions
at short distance, higher derivatives and/or higher-dimension
operators could be indispensable; (b) the Wilsonian matching of
effective field theories to QCD as discussed by Harada and
Yamawaki~\cite{HY:PR} requires that the parameters of the
effective theory be $intrinsically$ density- and
temperature-dependent. This would mean that the coupling constants
-- and not just the masses of the vector mesons -- will have
$intrinsic$ dependence which may not be fully accounted for by the
response to the background skyrmion matter that is taken into
account in our model. This feature was noted already at nuclear
matter density~\cite{song,BR03} where the $\omega$ coupling was
found to decrease along with the BR scaling mass of the $\omega$
in order to obtain the saturation and binding of nuclear matter;
(c) in describing dense matter approaching chiral restoration in
terms of $explicit$ (as opposed to integrated-out) massive degrees
of freedom, the lowest-lying vector mesons may not be sufficient.
It may be that the tower of vector mesons as suggested in the
``open moose" model of Son and Stephanov~\cite{son-stephanov} --
which is conjectured to be dual to QCD -- may have to be
implemented consistently with the symmetries of QCD. These points
are being investigated and will be reported in future
publications.


\section*{Acknowledgements}
The authors acknowledge helpful comments from Terry Goldman.
Vicente Vento is grateful for the hospitality extended to him by
KIAS, where his contribution to this investigation was carried
out. This work was partially supported by grants MCyT-BFM2001-0262
and GV01-26~(VV) and KOSEF Grant R01-1999-000-00017-0~(BYP).

\end{document}